\documentclass[prl,aps,twocolumn,floatfix,amsmath,amssymb,superscriptaddress,tightenlines]{revtex4}
\usepackage{graphicx}% Include figure files
\usepackage{epstopdf}
\newcommand{\be}{\begin{equation}}
\newcommand{\ee}{\end{equation}}
\usepackage{amsfonts}
\usepackage{bm}
\usepackage{color}
\begin{document}

\date{\today}
\title{Thermodynamic singularities in the entanglement entropy \\
at a 2D quantum critical point}
\author{Rajiv R. P. Singh}
\affiliation{Physics Department, University of California, Davis, CA, 95616}

\author{Roger G. Melko}
\affiliation{Department of Physics and Astronomy, University of Waterloo, Ontario, N2L 3G1, Canada}
\affiliation{Perimeter Institute for Theoretical Physics, Waterloo, Ontario N2L 2Y5, Canada}

\author{Jaan Oitmaa}
\affiliation{School of Physics, University of New South Wales, Sydney 2052, Australia}

\begin{abstract} 
We study the bipartite entanglement entropy of the two-dimensional (2D) transverse-field 
Ising model in the thermodynamic limit using series expansion methods. 
Expansions are developed for the Renyi entropy around both the small-field and large-field limits,
allowing the separate calculation of the entanglement associated with
lines and corners at the boundary between sub-systems.
Series extrapolations are used to extract subleading power laws and logarithmic
singularities as the quantum critical point is approached. 
In 1D, we find excellent agreement with exact results as well as quantum Monte
Carlo simulations.  In 2D, we find compelling evidence that the 
entanglement at a corner is significantly different from a free boson field theory.
These results demonstrate the power of the series expansion method for calculating
entanglement entropy in interacting systems, a fact that will be particularly useful
in future searches for exotic quantum criticality in models with and without the sign problem.
\end{abstract}

\maketitle

{\it Introduction:}
The study of entanglement properties of ground states of one-dimensional (1D) statistical systems and free field
theories in arbitrary dimensions is a very mature 
field \cite{cardy,peschel,ALreview}.
Many exact results have been established,
and numerical methods such as the Density Matrix Renormalization Group (DMRG) enable studies  of relatively
large system sizes in 1D \cite{DMRG}. In contrast, the study of entanglement properties for ground states of
interacting quantum lattice models in higher dimensions
is a subject still in its infancy \cite{max,senthil,max2,moore}.
In particular, although a great potential exists to connect properties of entanglement
to universality at quantum critical points (QCPs) \cite{QCscaling}, the critical scaling behaviors
of very few interacting lattice models are known.
Ultimately, the study of entanglement entropies may provide unique signatures of novel or deconfined QCPs \cite{DQCP}. Yet, much work is required before this advance is possible;
little is quantitatively known about the
nature of the singularities and crossovers at a QCP as a function of system size
and other thermodynamic parameters.

Recent developments in Quantum Monte Carlo methods offer a promising avenue for calculating entanglement
properties of higher dimensional quantum lattice models \cite{Hastings10,XXZ}. Another fruitful approach is the study of
entanglement in suitably parameterized variational wavefunctions \cite{Zhang1,Zhang2}. DMRG and Matrix Product State methods provide other
powerful variational approaches to study quantum entanglement in higher dimensional systems \cite{verstrate-wen, TEV, yao-balents}.
However, in contrast to these methods that require careful scaling analyses of finite-size lattices,
series expansions at $T=0$ provide a simple yet powerful alternative approach
to studying ground state entanglement entropy directly in the thermodynamic
limit. Calculations are carried out order-by-order in perturbation 
theory as a power series in some expansion variable $\lambda$, providing a
pedagogically transparent introduction to the development of entanglement entropy in many-body systems.
These expansions are typically convergent inside a phase, but become singular as a phase boundary is approached.
Once the expansions are developed to some order (in practice typically of order $10$), series extrapolation methods
can be used to approximate the singular behavior in entanglement 
near a QCP. 

Here, we use series expansions to calculate the thermodynamic singularities of
the 2D quantum critical point in the
transverse-field Ising model \cite{book},
\begin{equation}
{\cal H}=-J \sum_{\langle i,j \rangle} \sigma_i^z \sigma_j^z-h\sum_i \sigma_i^x,
\end{equation}
where the first sum runs over the nearest-neighbor bonds of the square-lattice and the second over
its sites. 
Both the limits $h=0$ and $J=0$ have very simple ground states,
and series expansions can be separately developed in $h/J$ or $J/h$. At small $h$ one has two
ordered ground states and the system has spontaneously broken $Z_2$ symmetry (called the ``ordered'' phase).
In developing the series expansion in $h/J$, we pick one of the two ground states of the system
to expand around. The state at $h=0$ is a simple product state with all the spins pointing
along the $z$ axis. At large $h$ (the ``disordered'' phase), one also has 
a simple product ground state, where every spin points along the $x$-axis. 
Thus both at $h=0$ and at $J=0$ the ground states have no entanglement between any pair of sites.
A quantum critical
point intervenes between the ordered and disordered phases, which is known to be in the universality class
of the 3D classical Ising model \cite{book}.
Using series expansion, we provide accurate calculation of the thermodynamic singularities in the entanglement entropy
for this universality class, demonstrating in particular differences from Gaussian free-field universality.

{\it Series Expansion Methods for Renyi entropies:} From a computational point of view, the Renyi entropies \cite{renyi} are particularly convenient measures of 
bipartite entanglement.
If we divide our system into two parts $A$ and $B$, such that each spin belongs to either $A$ or $B$,
then the ground state of the full system can be written in the local basis as
\begin{equation}
|\Psi_g\rangle=\sum_a\sum_b \psi_{a,b} |a \rangle |b \rangle,
\end{equation} 
where $a$ and $b$ refer to basis states for subsystems $A$ and $B$ respectively.
The Renyi entropies are defined as:
\begin{equation}
S_n=\frac{1}{1-n} \ln \left[{ {\rm Tr}(\rho_A^n) }\right],
\end{equation}
where, the trace is over all the states of the subsystem $A$ and
the reduced density matrix for the subsystem $A$ is given by the matrix elements
\begin{equation}
\langle a_1| \rho_A |a_2 \rangle =\sum_b \psi_{a1,b}^*\psi_{a2,b}.
\end{equation}

\begin{figure} {
\includegraphics[width=1.2 in]{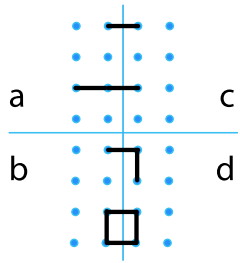} \caption{
The infinite square plane is partitioned into four quadrants
$a$, $b$, $c$ and $d$.
The region $A$ could be a half-plane such as $a \cup b$ or a
quadrant such as $b$ or $c$, while the rest of the square-lattice
forms the region $B$. For the former partition, several low-order
clusters that cross the boundary between $A$ and $B$ are also shown.
\label{planes}
}
} \end{figure}

In this paper we will focus attention solely on the second Renyi entropy $S_2$. We will divide the
infinite system into two subsystems such that the subsystem $A$ is either a half-plane or a quadrant
(See Fig.~1). We begin with the case when $A$ is a half-plane.
First non-zero terms in perturbation theory arise when pairs of spins from across the 
dividing line get entangled. Because the entropy is an extensive measure, each such pair contributes
equally to the sum and it leads to an entropy proportional to the length of the boundary. In the next
order either a pair of spins from one side can be entangled with one spin from the other side, or
a pair of spins from one side can entangle with a pair of spins from the other side. These contributions
have a natural graphical interpretation in terms of clusters that go across the boundary separating
$A$ and $B$ (See Fig.~1). 
The linked cluster method \cite{book,series-review} allows one to separate the entanglement that comes 
from a pair of spins versus
the additional entanglement that comes from a larger cluster of spins. Using the principle of inclusion 
and exclusion one can find the additional entanglement from a larger cluster 
(also called the weight of the cluster $W$)
by calculating the full entanglement for that cluster of spins when the perturbations are turned on, and then
subtracting from it the weight of all its subclusters.
\begin{equation}
W(c)=S_2(c)-\sum_{s} W(s),
\end{equation}
where the sum is over all subclusters of the cluster $c$. In the thermodynamic
limit, 
one can use the translational symmetry along the length of the boundary to write
the entropy per unit length as
\begin{equation}
s_2= S_2/L=\sum_{c_d} W(c_d).
\end{equation}
Here the sum is over all translationally distinct clusters $\{c_d\}$.
Expanding as a power series in $\lambda=h/J$ or $\lambda=J/h$ one obtains
\begin{equation}
s_2=\sum_n p_n \lambda^n.
\end{equation}

To obtain the corner terms, we need to consider the case where $A$ is a quadrant (See Fig.~1).
In fact, by considering different choices for the division of the infinite lattice into $A$ and $B$ it
is possible to completely cancel out the line contributions \cite{ourprl}. 
If we calculate the entanglement entropy for
(i) when $A$ is the quadrant (b) and (ii) when $A$ is the quadrant (c), then 
their sum will amount to entanglement from
two $90$ degree corners plus two infinite lines that cut across the lattice. The line contributions can
be subtracted off by subtracting the entanglement entropies for the cases, where $A$ is the half plane
formed by (i) $a \cup b$ and (ii) $a \cup c$. This subtraction can be done on a graph by graph basis.
Thus series for a corner, in the thermodynamic limit, can be
expressed in terms of graphs that lie at the intersection of two lines. This leads to the expansions
for the entanglement entropy at a single corner as
\begin{equation}
c_2=\sum_n q_n \lambda^n.
\end{equation}
The coefficients $p_n$ and $q_n$ are calculated to order $14$ in $J/h$ and 
order $24$ in $h/J$ and provided in supplementary material \cite{supplement}.

Note that one of the advantages of the series expansion method is that the line and corner contributions are
obtained separately. In higher dimensions, entropy associated with each type of manifold, planes, lines, corners 
can also be calculated separately.

{\it Series analysis:}
It is clear from the formalism that the `area law' is built into the series expansion method,
namely that the entanglement entropy scales with the boundary `area' between 
subsystems.
As long as the perturbation theory converges, the `area law' continues to hold. This is consistent with general
arguments that gapped phases obey the area law \cite{ALreview}. As one approaches a quantum critical point, where the gap goes
to zero, the series become singular. One can study whether the area law continues to hold at the critical point, and 
the nature of the critical singularity, by analyzing the limiting behavior of the series using extrapolation methods.

Series extrapolations can deal with convergent or divergent power-law singularities by using differential 
approximants \cite{baker,fisher}. One represents the function 
of interest $f(\lambda)$ by a differential equation of the form
\begin{equation}
Q_m(\lambda){df(\lambda)\over d\lambda}+P_n(\lambda) f(\lambda)=U_j(\lambda).
\end{equation}
Here $Q_m(\lambda)$, $P_n(\lambda)$ and $U_j(\lambda)$ are polynomials of order $m$, $n$ and $j$, which
are obtained such that the differential equation correctly reproduces the first $m+n+j+2$ powers in the
series expansions for the function $f(\lambda)$. 
The singularities of the function arise at $\lambda$ values where
$Q_m(\lambda)=0$. If the location of the critical point $\lambda_c$ 
is known by some other means, one can
put the additional constraint that $Q_m(\lambda_c)=0$. This is called biasing the critical point. One
can then study the resulting power-law singularity at $\lambda_c$. Note that the inhomogeneous term $U_j(\lambda)$
is essential to allow a non-zero slowly varying background in addition to a power-law singularity.

In the special case of a log singularity, one can first differentiate the function $f(\lambda)$, 
with respect to $\lambda$, and then 
represent it by a ratio of polynomials (also called Pad\'e approximants),
${df(\lambda)\over d\lambda}={P_n(\lambda)\over Q_m(\lambda)}.$
The log singularities arise for $f(\lambda)$ 
where $Q_m(\lambda)=0$. 

\begin{figure} {
\includegraphics[width=2.7 in]{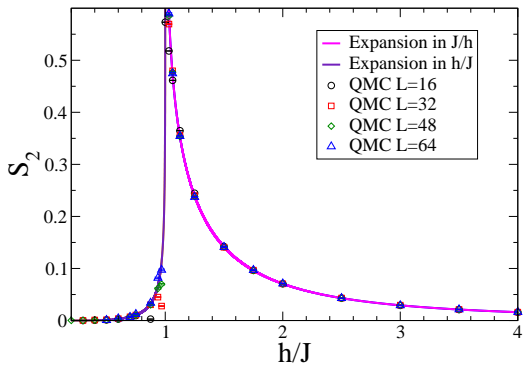} \caption{
Entanglement entropy of the transverse field Ising chain, for two edges, obtained by series expansions. For comparison QMC data on finite systems are also shown.
In the ordered phase $\log{2}$ has been subtracted from the QMC data to correspond to
the fact that series expansions are done around a single ordered ground state.
\label{1d}
}
} \end{figure}

First, we discuss the results for the transverse-field Ising chain for which closed form expressions for
the von Neumann entropy and asymptotic expressions for the Renyi entropies are available in the
literature \cite{peschel}. Our goal here is not to study the 1D model per se, but to simply see how well
one can estimate the critical properties using series expansions of the same length that can be
done in any dimension. In Fig.~2, the results of series extrapolation are shown. 
As a comparison, we also show finite-size data from Quantum Monte Carlo (QMC) on an $L$ length chain with periodic boundary
conditions, where $A$ and $B$ are both of length $L/2$.
The QMC was performed using a $T=0$ projector method 
with cluster updates \cite{AndersTFIM}, adapted to calculate Renyi entropy via the {\it Swap} operator \cite{Hastings10} on a replicated system \cite{Stephen}. 
The comparison shows that both the series expansion and QMC results are
very accurate, at least until one gets very close to the critical point, where finite size effects
become large and the QMC data drops away from the series extrapolation curve.

In 1D, the boundary `area' between subsystems is just a point or a corner. At the critical point this
corner contribution diverges logarithmically leading to a breakdown of the `area law'. The second Renyi
entropy associated with a single boundary is given asymptotically close to the critical
point by the expression \cite{cardy,peschel},
\begin{equation}
S_2={c\over 8}\log{\xi},
\end{equation}
where the central charge $c={1\over 2}$ for this model and the correlation length $\xi$ diverges as 
$1/|1 -\lambda|$
as $\lambda$ approaches unity. This means that the coefficient of the logarithmic singularity in $\log{|1-\lambda|}$
should equal $-0.0625$. Series extrapolations (with the length of series comparable to what is calculated in 2D)
give different answers from the expansions in $h/J$ and $J/h$. From one side one estimates the coefficient of the log
singularity to be $-0.053(1)$ where as from the other side one obtains $-0.077(2)$. The internal consistency
of the approximants leads to an unusually small estimate of the systematic uncertainty. In fact, one finds that
with increasing order both terms are changing in the right direction but only by about $0.001$ and $0.002$
respectively in each order. 
If we make the reasonable assumption that the coefficient must be the same from both sides and thus average
the two answers, one would obtain $-0.065 (12)$, which gives a stringent
limit on the uncertainty in the calculations.  

We now turn to the 2D transverse-field Ising model.
The line and corner terms for the entanglement entropy of the 
2D transverse-field Ising model are shown in Fig.~3.
In this case, the critical point is biased to
the value of $h/J=3.044$, a value determined previously \cite{book,Reiger}. 
Clearly the line term has a sharp peak at the critical point,
but it does not diverge, implying that the area law continues to
hold at the critical point. One expects the power-law singularity
to be of the form $|\lambda-\lambda_c|^\nu$, where $\nu$ is the critical
exponent for the divergence of correlation length in the $3D$ classical
Ising model. Our series extrapolations lead to estimates for $\nu=0.60 (2)$
and entropy per unit length at the critical point of $s_2^c=0.0324 (3)$ from one side
and $\nu=0.66 (3)$ and $s_2^c=0.0350 (3)$
from the other side. Averaging these we get, $\nu=0.63 (3)$ and
$s_2^c=0.337 (13)$. These values are clearly consistent with the known value
of $\nu=0.629 (2)$ \cite{ising3d} and recent QMC estimate of $s_2^c=0.0332 (4)$ \cite{preprint}.

The corner term is biased to have a log singularity. We estimate the
coefficient of the logarithm to be $0.0059 (3)$ from one side and $0.0077 (1)$
from the other side. Averaging the two, we get $0.0068 (9)$.
One can convert the logarithm in the variable $|\lambda_c-\lambda|$ into $\log(\xi)$
by dividing by $-\nu$, where $\nu$ is the correlation length exponent. Thus, we
estimate, asymptotically, for a single corner
\begin{equation}
c_2=(-0.011\pm 0.001) \log(\xi).
\end{equation}
These results are clearly distinct from the free field theory result of
Casini and Huerta who obtain $c_2=-0.0062$ \cite{casini}.
The QMC in Ref.~\cite{preprint} quotes a value of $ 4c_2=-0.03\pm 0.01$,
with large uncertainties that can not be distinguished from
free field theory.  

\begin{figure} {
\includegraphics[width=2.7 in]{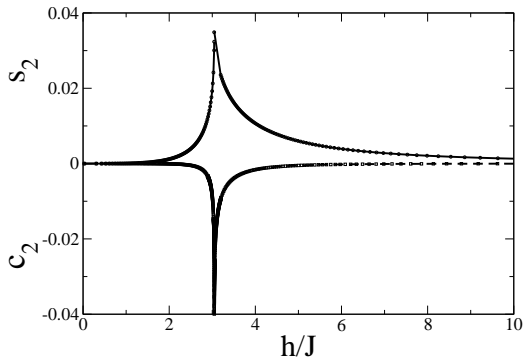} \caption{
`Area law' term $s_2$ and corner term $c_2$ of the entanglement entropy of the 2D transverse-field 
Ising model obtained by series expansions
in the variables $h/J$ and $J/h$. In each case two approximants with critical
point biased at $h/J=3.044$ are shown.
\label{2d-line}
}
} \end{figure}

{\it Discussions:}
We have shown that series expansions can be used to calculate thermodynamic singularities in the
entanglement entropy at 2D quantum critical points (QCPs) with about $10$ percent accuracy.
We have provided compelling evidence that the entanglement entropy produced at a corner
in the boundary between subregions in the 2D transverse field Ising model QCP is different from 
that predicted in a free boson field theory  \cite{casini}.  Indeed, the transverse field Ising model QCP is
the classical 3D Ising universality class, which is distinct from the free (Gaussian) universality class.
Currently there are no theory predictions for the corner log for the transverse field Ising model.

Through our calculations of $c_2$ (and confirmed by our accurate estimate of the exponent $\nu$), we have
 demonstrated that series expansions already suffice to distinguish between different 
universality classes \cite{casini,ann}.  It would be useful to study a range of models 
on different lattices to further consolidate the notion of universality. 
Given that the numerical values of critical parameters are largely unknown,
comparison between series expansions and QMC data, where available, would be most useful. 
Series expansions can also
be developed in higher than two dimensions and also for other Renyi indices $n$. 
These can help address questions related to upper critical dimensionality,
boundary correlation functions and possible singularities as a function of the Renyi index $n$ \cite{max2}.
For example, how does the logarithmic singularity associated with the corners depend on dimensionality, 
and does it have a simple limit above the upper critical dimension?

A weakness of the series method, as presented here, is the inability to study topological entanglement entropy.
The calculations discussed here are all related to the boundary between subsystems and in finite orders of perturbation
theory only the degrees of freedom at finite distance from the boundary get entangled. This can not
address topological entanglement, which is inherently long-ranged. 
It would be interesting to explore the possibility of addressing this through an approach involving 
degenerate perturbation theory \cite{schmidt}.

From a computational point of view, series expansion methods may be particularly useful in studying
interacting Fermion models and frustrated spin models. 
Quantum critical points are, in principle, accessible to high temperature
series expansions, which might provide a useful route to studying $t-J$ and Hubbard models. At $T=0$,
one should be able to look for exotic critical points 
at the boundary of magnetically ordered phases, 
or at the boundary between ordered phases and gapped spin liquids \cite{DQCP,IsakovXY}.
Indeed, several recent works \cite{verstrate-wen,yao-balents}
have argued that a spin liquid state may arise in the 
frustrated $J_1-J_2$ square-lattice Heisenberg model.
Investigations of the entanglement scaling at critical points
contained in this model will be pursued in future.

{\it Acknowledgments --}
The authors thank M.~Hastings, A.~Sandvik, S.~Inglis and M. Metlitski
for enlightening discussions.
RGM thanks the Boston University Condensed Matter Theory visitor's program for hospitality during a visit.
This work is supported by NSERC of Canada (RGM) and NSF grant No DMR-1004231 (RRPS). 
QMC Simulations were performed using the computing facilities of SHARCNET.

%\end{document}

%\begin{document}

\pagebreak

{\bf Supplementary Material for
``Thermodynamic singularities in the entanglement entropy at a 2D quantum critical point''
by R. R. P. Singh, R. G. Melko and J. Oitmaa}
 
%\maketitle

This supplement provides the reader with further technical data in support of the main part 
of the paper. First we present tables of series coefficients and then tables showing
estimates of the log singularity

\begin{table}[ht]
  \begin{tabular}{| l | c | }
  \hline
    n &  $p_n$     \\ \hline \hline
    2 & 0.125 \\ 
    4 & 0.4921875 \\ 
    6 & 2.38053385 \\ 
    8 & 14.0315145 \\ 
   10 & 90.7878208 \\ 
   12 & 625.42033 \\ 
   14 & 4501.17322 \\ \hline
 \end{tabular}
   \caption{\label{series-coeff1}
Series expansion coefficients for entanglement entropy per unit length $s_2$
($p_n$ in Eq.~7 in the paper)
for 2D Transverse Field Ising model in the
variable $J/h$.
   }
\end{table}

\begin{table}[ht]
  \begin{tabular}{| l | c | }
  \hline
    n &  $p_n$     \\ \hline \hline
    4 & 5.425349E-05 \\ 
    6 & 2.9669904E-06 \\ 
    8 & 3.61269111E-07 \\ 
   10 & 2.23023576E-08 \\ 
   12 & 2.64005223E-09 \\ 
   14 & 1.64460642E-10 \\ 
   16 & 2.27121391E-11 \\ 
   18 & 1.43038774E-12 \\ 
   20 & 1.84792391E-13 \\ 
   22 & 1.47160415E-14 \\ 
   24 & 1.56354826E-15 \\ \hline
 \end{tabular}
   \caption{\label{series-coeff2}
Series expansion coefficients for entanglement entropy per unit length $s_2$
($p_n$ in Eq.~7 in the paper)
for 2D Transverse Field Ising model in the
variable $h/J$.
   }
\end{table}

\begin{table} [ht]
  \begin{tabular}{| l | c | }
  \hline
    n &  $q_n$     \\ \hline \hline
    4 & -0.21875   \\ 
    6 & -1.578125  \\ 
    8 & -12.1454976 \\ 
   10 & -94.8888162 \\ 
   12 & -757.353415 \\ 
   14 & -6151.86971 \\ \hline
\end{tabular}
   \caption{\label{series-coeff3}
Series expansion coefficients for entanglement entropy of a corner $c_2$
($q_n$ in Eq.~8 in the paper)
for 2D Transverse Field Ising model in the
variable $J/h$.
   }
\end{table}

\pagebreak

\begin{table} [ht]
  \begin{tabular}{| l | c | }
  \hline
    n &  $q_n$     \\ \hline \hline
    6 & -4.23855252E-07 \\ 
    8 & -7.99438098E-08 \\ 
   10 & -9.17332192E-09 \\ 
   12 & -8.55136841E-10 \\ 
   14 & -1.02753121E-10 \\ 
   16 & -9.44954642E-12 \\ 
   18 & -1.09399979E-12 \\ 
   20 & -9.94186394E-14 \\ 
   22 & -1.13706509E-14 \\ 
   24 & -1.08159628E-15 \\ \hline
\end{tabular}
   \caption{\label{series-coeff4}
Series expansion coefficients for entanglement entropy of a corner $c_2$
($q_n$ in Eq.~8 in the paper)
for 2D Transverse Field Ising model in the
variable $h/J$.
   }
\end{table}

\begin{table} [ht]
  \begin{tabular}{| l | c | c| c| }
  \hline
    $(J/h)^2_c$ &  amplitude & m & n     \\ \hline \hline
  0.107922179 & 0.00762850877 & 2 & 4 \\
  0.107922179 & 0.00772136388 & 2 & 5 \\
  0.107922179 & 0.00754811631 & 3 & 3 \\
  0.107922179 & 0.00775324771 & 3 & 4 \\
  0.107922179 & 0.00761989088 & 4 & 2 \\
  0.107922179 & 0.00786590457 & 4 & 3 \\
  0.107922179 & 0.00777476393 & 5 & 2 \\ \hline
\end{tabular}
   \caption{\label{pade1}
Pade approximant estimates for amplitude of log-divergence
for 2D Transverse Field Ising model in the
variable $(J/h)^2$. The first column is the biased location of
the critical point. The second is the amplitude for the log divergence.
The third and fourth give $m$ $n$ values for the Pade approximants.
   }
\end{table}

\pagebreak

\begin{table} [ht]
  \begin{tabular}{| l | c | c| c| }
  \hline
    $(h/J)^2_c$ &  amplitude & m & n     \\ \hline \hline
  9.265936 & 0.00619029914 & 3 & 6 \\
  9.265936 & 0.00590459618 & 3 & 7 \\
  9.265936 & 0.00573168376 & 3 & 8 \\
  9.265936 & 0.00698086931 & 4 & 5 \\
  9.265936 & 0.00586117615 & 4 & 6 \\
  9.265936 & 0.0053144883  & 4 & 7 \\
  9.265936 & 0.00588231756 & 4 & 8 \\
  9.265936 & 0.00578324783 & 5 & 4 \\
  9.265936 & 0.00624218823 & 5 & 5 \\
  9.265936 & 0.00560957049 & 5 & 6 \\
  9.265936 & 0.00584200216 & 5 & 7 \\
  9.265936 & 0.00608591271 & 6 & 4 \\
  9.265936 & 0.00604662478 & 6 & 5 \\
  9.265936 & 0.00613351483 & 6 & 6 \\
  9.265936 & 0.00604144926 & 7 & 4 \\
  9.265936 & 0.00607802345 & 7 & 5 \\
  9.265936 & 0.00624832351 & 8 & 4 \\ \hline
\end{tabular}
   \caption{\label{pade2}
Pade approximant estimates for amplitude of log-divergence
for 2D Transverse Field Ising model in the
variable $(h/J)^2$. The first column is the biased location of
the critical point. The second is the amplitude for the log divergence.
The third and fourth give $m$ $n$ values for the Pade approximants.
   }
\end{table}

\end{document}